\documentstyle[psfig]{l-aa}
\begin{document}
\thesaurus {11.05.2, 11.07.1, 11.09.2, 11.16.1, 11.16.2, 11.19.2}

\title{ Statistics of optical warps in spiral disks }
\author{Vladimir Reshetnikov\inst{1,2}, Fran\c coise Combes\inst{2}}    

\offprints{F.~Combes \hfill\break(e-mail: bottaro@obspm.fr)}   

\institute{Astronomical Institute of St.Petersburg State University,     
   198904 St.Petersburg, Russia     
\and
  DEMIRM, Observatoire de Paris, 61 Av. de l'Observatoire,     
 F-75014 Paris, France}  

\date{Received 1998; accepted}

\maketitle
\markboth{V. Reshetnikov \& F. Combes:
Optical warps }{}

\begin{abstract}
We present a statistical study of optical warps in a sample of
540 galaxies, about five times larger than previous samples.
About 40\% of all late-type galaxies reveal S-shaped warping of
their planes in the outer parts. Given the geometrical parameters
and detection sensitivity, this result suggests that at least half of all
galaxy disks might be warped. 
We demonstrate through geometrical simulations that some apparent warps
could be due to spiral arms in a highly inclined galaxy. 
The simulations of non warped galaxies give an amount of
false warps of $\approx$ 15\%, while simulations of warped galaxies
suggest that no more than 20\% of the warps are missed.
We find a strong positive correlation of observed warps  with
environment, suggesting that tidal interaction have a large influence
in creating or re-enforcing warped deformations.
 

\end{abstract}

\section{Introduction}

Most spiral galaxies that can be traced in neutral hydrogen
exhibit a warped plane (e.g. Sancisi 1976; Bosma 1981; Briggs 1990,
Bottema 1995). Our own Galaxy is strongly warped from the solar
radius (e.g. Burton 1992, Diplas \& Savage 1991).

While the wide majority of warps have been detected in HI, outside
the optical disks, some optical warps exist (M31, Innanen et al 1982;
van der Kruit 1979, van der Kruit \& Searle 1981, 82). In our own
Galaxy, the stellar disk appears to follow the HI warp
(Porcel \& Battaner 1995, Reed 1996). The best
evidence for warped optical disks occurs in tidally interacting
galaxies (for instance UGC~3697, Burbidge et al 1967, or
NGC~4656, Weliachew et al 1978). Sanchez-Saavedra et al (1990)
have surveyed 86 edge-on galaxies in the northern sky, and claim
that about half of the galaxies display some sort of warping of
the plane. This high percentage would suggest that most galaxies
are warped, since projection effects must mask at least some fraction of
the warps, those with line of nodes perpendicular to the line of
sight (Sanchez-Saavedra et al estimate that the observed fraction
of warps should be multiplied by 1.7 for this).
Sanchez-Saavedra et al results were confirmed by Reshetnikov (1995)
from the study of a complete sample of 120 northern edge-on
spiral galaxies. 

Extended warps represent a dynamical puzzle, since they should be rapidly washed
out by differential precession, and end up in corrugated disks. Many 
models have been proposed for their persistence, from an intergalactic
magnetic field (Battaner et al 1990), discrete bending modes
(Sparke 1984, Sparke \& Casertano 1988), misaligned dark halos
(Dubinski \& Kuijken 1995), or cosmic infall and outer gas accretion
(e.g. Binney 1992). Another possibility is that warps are self-gravitating,
which reduces differential precession (Pfenniger et al 1994). In any case,
the bending of the plane above the equator defined by the inner galaxy
allows to explore the 3D-shape of dark matter, and to probe its potential.
 It is therefore of prime importance to tackle the warp formation 
mechanisms, and to enlarge the statistical data, especially on
optical warps (section 2 and 3). Optical warps are weaker than HI
warps and can be easily contaminated by dust or projected spiral arms,
when the galaxy is not exactly edge-on; we thus undertake simulations
of dusty spiral galaxies, with or without warps, to estimate
quantitatively the biases (section 4).

\section{Sample and measurements}

In order to find a large enough and unbiased sample 
for optical warps detection, we selected the Flat Galaxy Catalogue
by Karachentsev et al (1993) (FGC). This catalogue 
has been built from the Palomar Observatory Sky
Survey and the ESO/SERC survey and contains 4455 galaxies 
with a diameter larger than 40$\arcsec$ and major-to-minor
axis ratio $a/b\geq$7. The FGC covers about 56\% of the whole
sky and is about 80-90\% complete for the galaxies with
blue diameter larger 0.7$\arcmin$.

We decided to consider the Digitized Sky Surveys\footnote{The
Digitized Sky Surveys were produced at the Space Telescope
Science Institute under U.S. Government grant NAG W-2166} (DSS) images 
of the FGC galaxies. Due to a better quality of the photographic
emulsions used for the southern sky survey, the galaxies digitized
using the ESO/SERC films extend to a surface brightness level
slightly fainter than the galaxies measured on the POSS films. 
Therefore, we selected the Southern Extention of FGC (FGCE) for our 
study. Our final sample consists of $all$ FGCE galaxies with
blue angular diameter between 1$\arcmin$ and 3$\arcmin$ and
coordinates  ${\rm 0.^{h}0}~\leq~\alpha({\rm 1950})~\leq~{\rm 14.^{h}0}$,
$\delta({\rm 1950})~\leq~{\rm -17.^{o}5}$. The sample includes
540 galaxies, which is about five times larger than previously
studied samples by Sanchez-Saavedra et al (1990) and
Reshetnikov (1995).

We extracted images of all the sample objects from the DSS.
The size of each retrieved square area was ten blue diameters of 
the investigated galaxy. Thus, typical images were
$\rm (350-1000)^{2}$ pixels, each 1.$\arcsec$7$\times$1.$\arcsec$7.
Then, we reduced edge-on galaxy images in the MIDAS environment.
In total, 526 of 540 sample objects (97.4\%) have images suitable for
warp detection  (the images of the remaining 14 galaxies are too
faint and knotty).

For each object we constructed isophotal maps of full square area
around the galaxy (with size 10$\times$galaxy diameter) and
of the investigated galaxy only with the faintest contour 
corresponding to 2$\times\sigma$ of the sky level (in densities) 
near the object. Large-scale maps of full area around the
object were used for the study of galaxy environment.
According to environment, we separated our sample in three
subsamples: isolated galaxies (without companions with
angular diameter larger than 1/5 of the primary within 5
optical diameters of the investigated object), non-isolated
galaxies (with companions) and interacting galaxies (obviously
interacting systems with tails, bridges, envelopes etc.).

\begin{figure}
\psfig{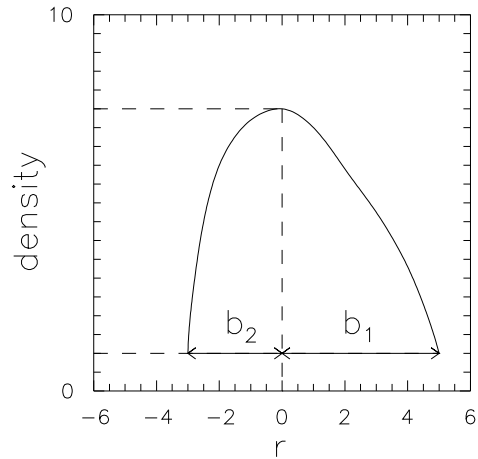}
\caption{Definition of the asymmetry index: $b_{1}/b_{2}$. The
solid line is a cut along the minor axis of the galaxy: the abcissa
is the distance along the minor axis, the ordinate is the observed
brightness density}
\label{definition}
\end{figure}

From the detailed map of each galaxy we measured the asymmetry index,
defined as the ratio  of distances measured perpendicular
to major axis from maximum intensity to outer isophote --
see fig \ref{definition}.
This index somewhat characterizes the
orientation of the disk relative to the line of sight together
with the amount of dust present in the disk. A larger
index corresponds, on average for a given dust content, to 
a less inclined galaxy. 

The identification of a weak optical warp in a galaxy disk
is not a simple procedure (for instance, in some cases
we could take for a warp a highly inclined spiral arm).
We identified a warp simply as a large-scale systematic
deviation of galaxy isophotes from the plane defined
by the inner ($\leq$1/2 of optical radius) region of a galaxy.
We fixed this plane as the  average position angle of
elliptically averaged isophotes; the galaxy center is defined as
the mean center of averaged isophotes within $\leq$1/2 of the optical radius
of the galaxy.
As a measure of warp we tried to use the difference
between position angles of elliptically averaged isophotes 
of central and outer regions of a galaxy. We found that
such an approach leads to a strong underestimate of the actual warp
angle. Therefore, we also chose the less objective but straightforward
procedure of eyeball estimation of the warp angle from
detailed isophotal maps of galaxies. 
After some experiments
we found that for warp angles ($\psi$ -  angle measured
from the galaxy centre, between the plane and average line from centre
to tips of outer isophotes)
larger than 2$^{\rm o}$-2.$^{\rm o}$5 such a procedure gives quite
reliable results. Our warp measurements refer to outer
regions of galaxies with estimated surface brightness level
$\mu_{B}~\sim~\rm 25.5$. Due to slightly varying (from field to field)
quality and depth of the digitized photographic films one can give 
only such general estimation of the surface brightness level.

\begin{figure}
\psfig{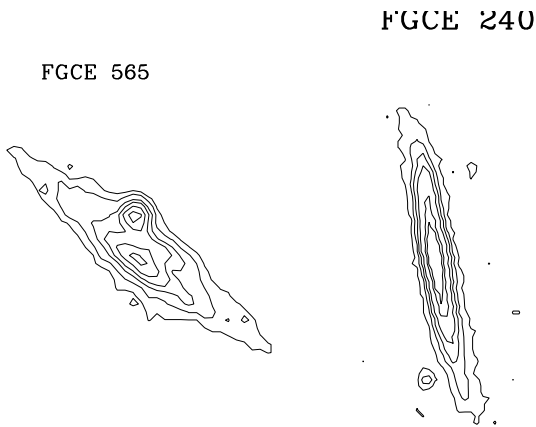}
\caption{ Example of a U-shaped warp: FGCE 565, and S-shaped warp: FGCE 240}
\label{examples}
\end{figure}

Two types of perturbation of the outer isophotes are distinguished:
the S-shaped warp, where the plane of the galaxy takes the shape
of an integral sign, i.e. rises on one side, and symmetrically
declines on the other. An example is shown in fig \ref{examples}.
The other type is an U-shaped warp, where the two sides rise together
(see fig \ref{examples}). This last deformation is linked in general
to a large asymmetry parameter, i.e. means that dust is hiding the far side
of the galaxy.
We plan to publish isophotal maps
of all S-type warped galaxies with 
$\psi~\geq~$4$^{\rm o}$ in a forthcoming paper ($\approx$
60 objects).

\section{Results}

The general results of our analysis are presented in Table 1.

About a quarter of the sample objects are classified as
isolated galaxies (without significant neighbours within 5 optical
diameters). This fraction depends on the angular size of galaxies:
18\%$\pm$3\% of the galaxies with angular diameter 
between 1.$\arcmin$0 and 1.$\arcmin$2, and 
34\%$\pm$6\% of those with diameters between 1.$\arcmin$7 and
3.$\arcmin$0 are found isolated. This correlation was also
found by Karachentseva (1973) as a function of apparent
magnitude: the fraction of isolated galaxies is higher for
brighter objects. This is certainly related to the isolation 
criterium itself, since the companion should be at
least 20\% the size of the primary, and large galaxies are
rarer.

The relative fraction of interacting galaxies -
6\% - is compatible with independent earlier estimations (for example,
Arp \& Madore 1977, Dostal 1979) based on the presence
of morphological signs of interaction. 

\begin{table*}
\caption[1]{Statistics of the sample galaxies (all numbers in \%)}
\begin{tabular}{llll}
\\ \\
\hline
\\
   & Isolated galaxies & Galaxies with companions & Interacting galaxies \\ 
   & (N=133)           & (N=360)                  & (N=33)               \\ \\  
\hline \\
Fraction        & 25.3$\pm$2.2 & 68.4$\pm$3.6 & 6.3$\pm$1.1  \\ \\
Warp type: \\ \\
\hspace*{0.5cm}without warp (all) & {\it 42$\pm$6} & {\it 24$\pm$3} & 
{\it 12$\pm$6}     \\
\hspace*{1.7cm}1.$\arcmin$0-1.$\arcmin$2 & 43$\pm$11 & 25$\pm$4 &
20$\pm$12 \\
\hspace*{1.7cm}1.$\arcmin$2-1.$\arcmin$7 & 35$\pm$8  & 28$\pm$4 & 0        \\
\hspace*{1.7cm}1.$\arcmin$7-3.$\arcmin$0 & 51$\pm$12 & 14$\pm$5 & 0 \\ \\
\hspace*{0.5cm}U-shaped (all) & {\it 33$\pm$5}  & {\it 36$\pm$3} & 
{\it 36$\pm$10} \\ 
\hspace*{1.7cm}1.$\arcmin$0-1.$\arcmin$2 & 38$\pm$10 & 37$\pm$5 &
33$\pm$15 \\
\hspace*{1.7cm}1.$\arcmin$2-1.$\arcmin$7 & 37$\pm$8  & 38$\pm$5 &
44$\pm$17 \\
\hspace*{1.7cm}1.$\arcmin$7-3.$\arcmin$0 & 24$\pm$8  & 29$\pm$6 & 0  \\ \\
\hspace*{0.5cm}S-shaped (all) & {\it 21$\pm$4} & {\it 36$\pm$3} & 
{\it 45$\pm$12} \\ 
\hspace*{1.7cm}1.$\arcmin$0-1.$\arcmin$2 & 19$\pm$7 & 34$\pm$5  & 40$\pm$16 \\
\hspace*{1.7cm}1.$\arcmin$2-1.$\arcmin$7 & 22$\pm$6 & 32$\pm$5 & 50$\pm$18 \\
\hspace*{1.7cm}1.$\arcmin$7-3.$\arcmin$0 & 22$\pm$8 & 52$\pm$9 & 67$\pm$47 \\ \\
\hline
\end{tabular}
\end{table*}

\begin{figure}
\psfig{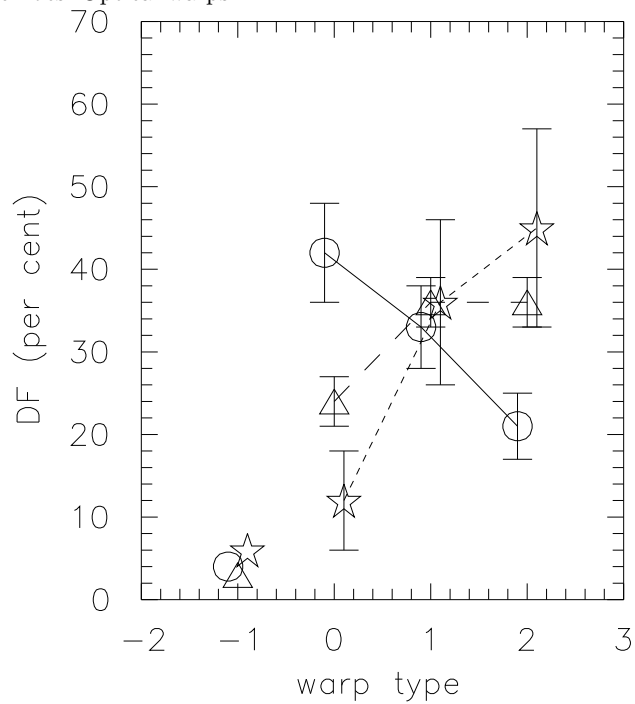}
\caption{Dependence of the detection fraction of isolated (circles,
solid line), non-isolated (triangles, dashed line) and
interacting (stars, short-dashed line) galaxies on the warp
type: -1 means uncertain type of warp due to overlapping of
the galaxies, stars projection etc; 
0 $-$ without warp; 1 $-$ U-shaped warp; 2 $-$
S-shaped warp}
\label{w-isol}
\end{figure}

As can be seen in Table 1 and fig. \ref{w-isol}, a significant fraction of 
isolated objects (40-50\%) does not show measurable warps while
among interacting galaxies non-warped objects are relatively rare -
10-20\%. On the contrary, the relative fraction of S-shaped
warps among isolated galaxies is about 20\% only, while among
interacting galaxies such objects constitute 40-50\%. 
Non-isolated objects (with close companions but without
obvious signs of interaction) are intermediate between
isolated and interacting galaxies. 
The relative fraction of U-shaped warps is the same - 30-40\% -
for all considered subsamples of galaxies.     

In order to check possible observational selection biases leading to
an easier detection of warps among nearby and large galaxies, 
we divided our sample into three parts, according to the
angular diameters: from 1.$\arcmin$0 to 1.$\arcmin$2,  
from 1.$\arcmin$2 to 1.$\arcmin$7, and from 1.$\arcmin$7 to
3.$\arcmin$0. All these samples are of comparable volumes. 
It is evident in Table 1 that the general observational trends 
described above do not depend on the angular size of the galaxy.  

\begin{figure}
\psfig{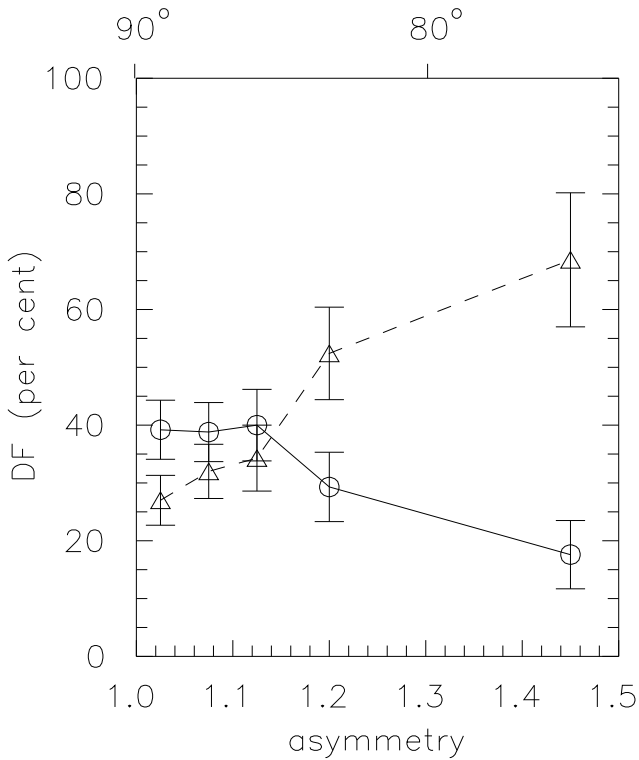}
\caption{Dependence of the detection fraction of U-shaped 
(triangles, dashed line) and S-shaped (circles, solid line) 
warps on the asymmetry index.}
\label{asym}
\end{figure}

Fig. \ref{asym}  presents the relation between the relative fraction
of U- and S-shaped warps and the asymmetry
index. This index correlates statistically with
galaxy inclination - larger values correspond to less
inclined disks. As one can see, the detection fraction of
S-shaped warps is constant (about 40\%) for edge-on
galaxies and decreases significantly with decreasing 
inclination. This reflects obviously an observational bias
- difficulty to detect a small optical warp for
non edge-on galaxy. On the contrary, the relative fraction
of galaxies with U-shaped warps increases systematically
with decreasing inclination. We can speculate, therefore,
that the U-shaped appearance can be related in many cases
to non edge-on orientation (for instance, due to dust
influence on the brightness distribution). Anyway, it
is evident in fig. \ref{asym} that among edge-on warped galaxies 
dominate objects with S-shaped warps.

\begin{figure}
\psfig{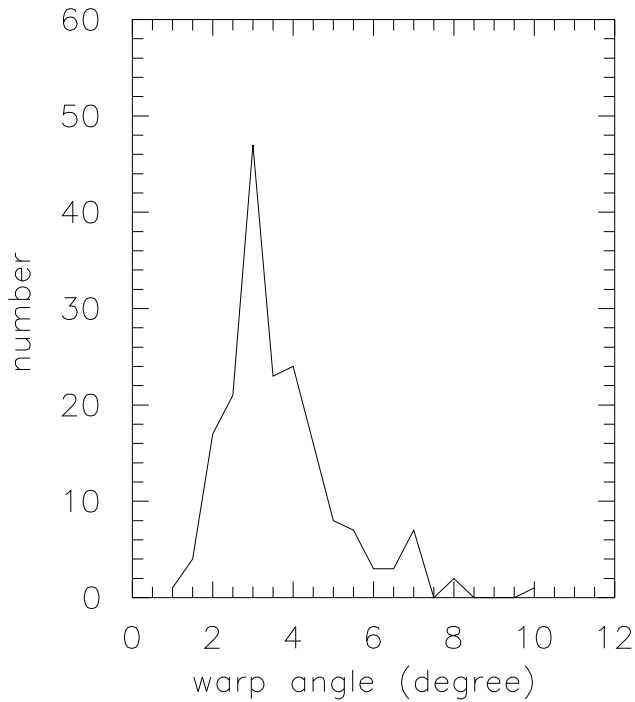}
\caption[]{Distribution of the observed warp angles for 
the galaxies with S-shaped warps.}
\label{dis-an}
\end{figure}

The distribution of observed warp angles is peaked at
$\psi~=~3^{\rm o}$ (fig. \ref{dis-an}). This value is somewhat smaller
than reported by Reshetnikov (1995) ($\rm 4^{o}-5^{o}$).
This difference is quite natural since the present study
is based on less deep (in the sense of surface brightness
level) material. The right wing of fig. \ref{dis-an}
distribution (for $\psi~>~\rm 3^{o}$) may be approximated
by $\propto\psi^{-5}$ law.
The observed distribution shown in fig. \ref{dis-an}
is strongly affected by our detection limit
($\rm 2^{o}-2.^{o}5$) for the angles $\psi~\leq~\rm 3^{o}$.
We detected small warps with $\psi~\leq~\rm 3^{o}$ 
in very thin, symmetric and nearly edge-on galaxies only.

\begin{figure}
\psfig{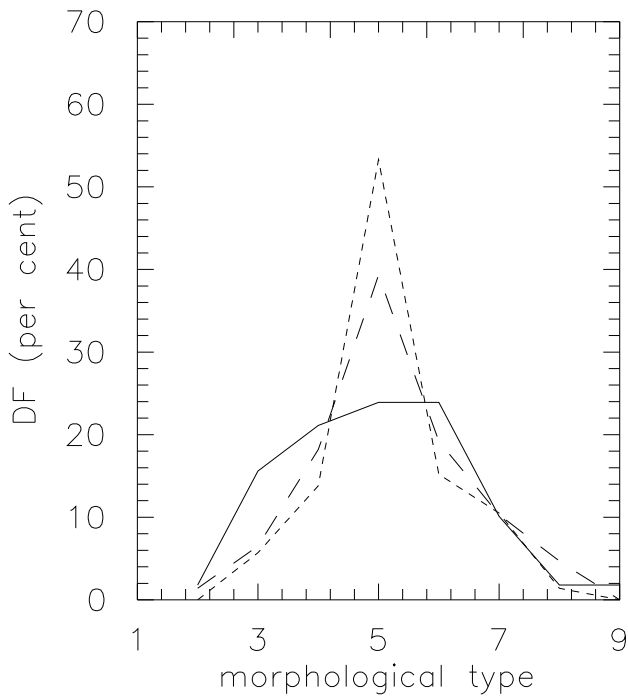}
\caption[]{Distribution of the sample galaxies on the morphological
type. Short-dashed line presents the galaxies with angular diameter
between 1.$\arcmin$0 and 1.$\arcmin$2, dashed line - 
1.$\arcmin$2-1.$\arcmin$7, and solid line - 1.$\arcmin$7-3.$\arcmin$0.}
\label{dis-typ}
\end{figure}

Fig. \ref{dis-typ} presents the distribution of the sample objects
according to morphological type given by Karachentsev et al (1993). 
The figure demonstrates
clearly that the morphological classification in the FGCE
depends strongly on the angular size of the galaxy:  
larger galaxies show a flatter distribution, while smaller 
concentrate near Sc (T$=$5) type. This is a natural
observational bias: a detailed classification is much easier
for larger objects.

\begin{figure}
\psfig{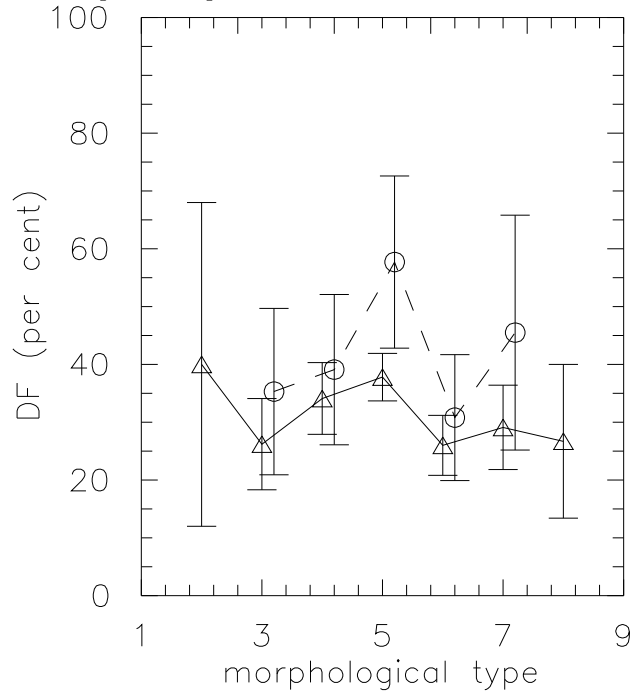}
\caption[]{Dependence of the observed detection fraction of
S-shaped warps on morphological type. 
Triangles - all the galaxies, circles - objects with diameters
within 1.$\arcmin$7-3.$\arcmin$0.}
\label{morph}
\end{figure}

In fig. \ref{morph} we plot the relative fraction
of S-shaped galaxies as a function of their morphological type.
The present statistics do not show
any significant correlation between the frequency of warps
and the morphological type.  

De Grijs (1997) has recently made a deep study of edge-on galaxies,
and on a sample of 44 galaxies determined a warp fraction of 64\%.
 This appears somewhat higher than the previous estimations, which 
could be due to several factors. First his study was made on deeper images,
down to blue surface brightness of 27 mag $arcsec^{-2}$ (compared 
to 25 mag $arcsec^{-2}$ for the other surveys), but this does not seem
the essential point, since his detected warps begin at a surface
brightness level of $\mu_B$ between 20 and 24 mag $arcsec^{-2}$.
Second, his sample does not contain only very flat galaxies,
since the selection criterion is the blue axis ratio $a/b$ at
least 3.1. Although the cataloged inclination is high ($i > 87^\circ$),
the latter is not well determined; the simulations of the next section
have shown that the apparent axis ratio can vary from 9 to 4, at a given
inclination between 80 and 85$^\circ$, only through variation of the 
position angle of the spiral arms with respect to the line of sight.
 For galaxies that are not very flat, the risk to find artificial
warps is high (cf fig \ref{stat-warp}).

To compare our estimation of the warp fraction to that of de Grijs (1997),
we have tried to repeat our study on his sample, chosing only the 
DSS isophotes, as we described in section 2. We find indeed a higher
fraction of warps than in the present sample:
7 U-shaped, and 18 S-shaped, i.e. a total fraction of 56\%.
One of the reason 
could be small number statistics, and another the selection 
criteria, in particular concerning the axis ratio.
Note that de Grijs (1997) selected non-interacting and unperturbed
galaxies, so that no conclusion about environment can be drawn.
 To better disentangle the effects of dust and spiral features
on determined warps, it is interesting to examine the near-infrared
images. The latter, however, often do not go as far out in radius
as the B-band images; but when it is possible to compare, 
there is no warp in the K-band (cf Grijs 1997). This is related
to the statistical results of Sanchez-Saavedra et al (1990) who
find almost half of the edge-on galaxies warped in the blue POSS 
sky survey, while only one third of them is warped in the red POSS plates
(see also Florido et al 1991).

Summarizing the results of three independent surveys of optical
warps in galaxies (Sanchez-Saavedra et al 1990, Reshetnikov 1995,
and the present study) one can make the following conclusions:

$\bullet$ About 40\% of spiral galaxies with $\rm T~\geq~2$ 
reveal S-shaped optical warps in their outer parts
($\mu_{B}~\geq~\rm 25$)
with typical amplitudes $\geq \rm 2^{\rm o}$. 
This high {\it observable} percentage would suggest that 
a large fraction, more than a half, of all spiral galaxies are warped.

$\bullet$ The probability of optical warp detection
does not depend on galaxy morphology (for objects later than Sab).

$\bullet$ Disks of more massive, large and luminous galaxies
are somewhat less warped (from a complete sample of 120
northern spirals with known magnitudes, diameters and maximum rotational 
velocities according to Reshetnikov 1995).

$\bullet$ The detection fraction of S-shaped warps depends
on galaxy environment: warped galaxies dominate among interacting
galaxies, are very frequent among galaxies with close companions,
and relatively rare among isolated objects.

\section{Simulations of projection effects}  

Artefacts due to the nearly edge-on projection of a galactic disk, which is not
homogeneous, but with spiral structure and related dust lanes, 
could in some cases bias the observations, and perturb our statistics on
warps; in this section we study these effects, to better
subtract the perturbations.

We are particularly interested in the two main projection effects:
first, if the galaxy disk is not exactly inclined by $90^\circ$, 
obscuration by dust in the plane affects more the far side than the
near side of both bulge and disk. The integrated light maxima are
shifted towards the near side, and the isophotes are off-centered
especially on the minor axis, since the importance of dust decreases
from the center to the outer parts of galaxies. This could mimic
a slight bend of the disk in the U-shape type;
second, if there exists a somewhat symmetric and contrasted $m=2$ spiral 
structure in the disk, which is not exactly edge-on, we can confuse the
spiral structure for the disk itself, and "see" the S-shape of the spiral
which happens actually to wind in an un-warped disk.

To simulate these effects, we build realistic spiral disks with
embedded dust, and reproduce their apparent isophotes as a function
of position angle and inclination. The radiative transfer is very simple,
including only absorption and no scattering, to give an idea
of the first order effets. Scattering has been taken into account
in previous modelisations, in order to find clear diagnostics of the
dust content of spiral disks (e.g. Byun et al 1994 and references
therein). It has been shown that scattering reduces the amount
of apparent extinction, and the effect is more important for face-on
galaxies, while it is not significant for edge-on disks that
concern us here. On the contrary, the effect of spiral structure
has not been investigated, and it is of primordial importance here.

\subsection{Disk structure}  

The overall global distribution in radius $r$ and height $z$ of the light
corresponds to an exponential stellar disk
$$
\rho_s(r, z) = \rho_s(0,0) exp ( -r/rs - |z|/hs )
$$
and a spherical Plummer bulge:
$$
\rho_b(r, z) = \rho_b(0,0) ( 1 + r^2/ab^2 ) ^{-5/2}
$$
The dust is only distributed in the disk, with the same overall
distribution, but different values for scale-length and scale-height, 
$rd$ and $hd$.
The density in the disk is then multiplied by the spiral 
function:
$$
f(\theta) = 1  + \Sigma_m f_m cos (m\theta -\phi_s)
$$
with the possible harmonics $m =$ 2 and 4 only; the phase is chosen 
to give a logarithmic spiral, outside of the radius $R_{bar}$
$$
\phi_s = \alpha \, log(r/R_{bar})
$$ 
with $R_{bar}$ = 4kpc (inside the disk is axisymmetric) 
and the dust spiral has the same shape, with a phase shift, i.e.
$\phi_d = \phi_s + \delta\phi$, with $|\delta\phi|$ =10$^\circ$, to account for 
the observation that the dust lanes in general lead/trail the stellar spirals.

To explore the influence of the bulge-to-disk ratio, we simulated
3 values of the mass ratio $M_b/M_d$ = 0, 0.1 and 0.3. The latter value
is already large for the samples we are considering (since we selected
very flat galaxies, with axis ratios $b/a$ lower than 0.15, and in 
average 0.11), but they are instructive and necessary to complete 
the statistics.
We adopt here a constant ratio of 4 between the disk and bulge 
scale length, i.e. a Plummer parameter $ab$ = 0.25 $rs$.
The scales of the stellar and dust disk are $rs$ =6 kpc, $hs$ =450pc,
$rd$ = 7 kpc, and $hd$ = 200pc.

The model galaxies were inclined on the sky to be nearly edge-on,
with 5 inclination angles $i =$ 80., 82.5, 85., 87.5 and 90$^\circ$.
To estimate the projection effects of the spiral arms, the
galaxies were also rotated around their rotation axis, through
9 position angles (from PA = 0 to 160 by 20$^\circ$).
We finally considered three values of the total dust optical depth
at $z=0$ for an edge-on galaxy, $\tau_0$ = 1, 5 and 10. Taking into account
the 3 bulge-to-disk ratios, 
these tests resulted in 405 galaxy models, from which we
estimate statistically the possible projection biases.

\begin{figure}
\psfig{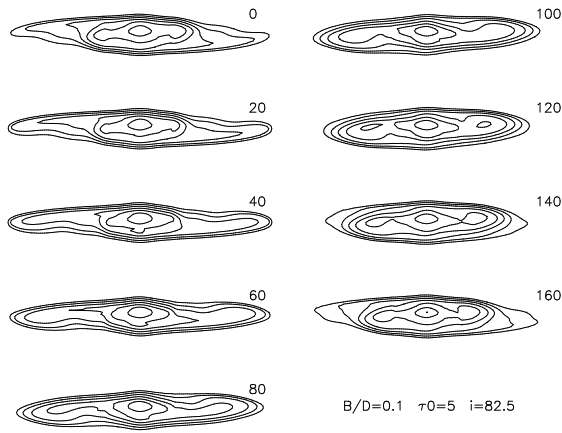}
\caption{ Some typical results from the simulations: logarithmic contours
of inclined ($i= 82.5^\circ$) galaxies, with bulge-to-disk ratio $B/D = 
0.1$, and a total edge-on optical depth of $\tau_0=5$. The position angle $PA$
is indicated in each frame at top right in degrees. }
\label{b1t5i82}
\end{figure}

\subsection{Absorption calculations}

Once the stellar density $\rho_s(r,\theta,z)$ and
dust density $\rho_d(r,\theta,z)$ are settled, the cube representing
the galaxy is rotated to the given orientation (PA, $i$),
and the light for any line of sight integrated, taken into account
progressively the absorption along the line of sight $s$ as
$$
I \propto \int ds \rho_s(s) exp (- \tau(s))
$$
with
$$
\tau(s) \propto \int_0^s ds' \rho_d(s')
$$

\begin{figure}
\psfig{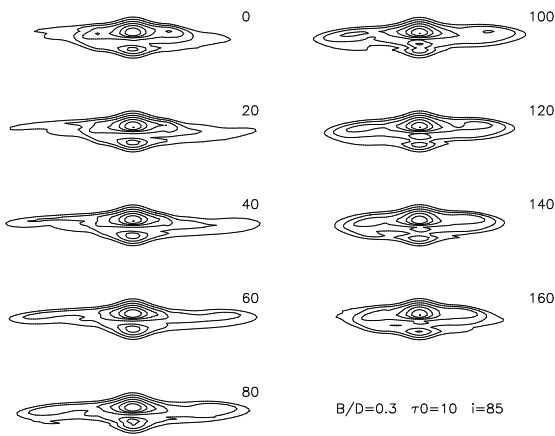}
\caption{ Same as previous figure, for more
inclined ($i= 85^\circ$) galaxies, with bulge-to-disk ratio $B/D = 0.3$,
and a total edge-on optical depth of $\tau_0=10$.}
\label{b3t10i85}
\end{figure}

Some results are shown in fig \ref{b1t5i82} for an inclination not exactly
edge-on. We can see how projection effects 
are producing S-shape warps through special viewing of the spiral
structure. The effect of dust is revealed in fig \ref{b3t10i85},
where U-shape warps are produced by asymmetric absorption.

\subsection{Statistical results}

\begin{figure}
\psfig{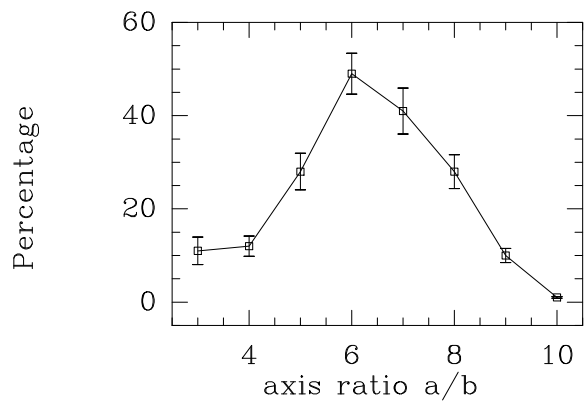}
\caption{ Percentage of apparent S-shaped warps, only due to
projection effects, as a function of apparent axis ratio for the model
galaxies. The error bars are the standard dispersion in N$^{-1/2}$. }
\label{stat-warp}
\end{figure}

From the 405 models computed, we keep only those flat enough to
be compared to galaxies in our sample, i.e. with axis ratio $a/b > 7$.
This is not equivalent to a constrain on inclination only, since it depends
on the orientation of the spiral structure with respect to the 
line of sight, as shown in fig \ref{b1t5i82}. 
 The main result is that projection effects can indeed produce false
S-shaped warps in about 50\% of cases, but only at low inclination,
$i \le 82^\circ$. At $i \sim 85^\circ$, the false S-shaped warps
are around 30\%, and they fall to 0\% at $i \approx 90^\circ$.
It is possible that these figures are overestimating the artefacts,
since all model galaxies had a nice contrasted spiral structure,
which is not the case in actual galaxies.  Our sample of
flat galaxies have $a/b > 7$, and average $b/a$ = 0.11. From the detailed
distribution of axis ratios in the observed sample, together with
figure \ref{stat-warp}, we estimate the percentage of false
S-shaped warps to 15\%.

We also searched for mocked U-shaped warps in those simulations. 
There were very little, of the order of 6\%, although this is subjective. 
One of the reason is that false U-shaped warps are expected
to be due to less inclined dusty galaxies, and in our sample there are
only nearly edge-on systems. Secondly, our model galaxies are always symmetrical,
and there must be in the observations many U-shape due to intrinsic
asymmetries. This is re-inforced by the spiral arm contrast which
is constant in our model galaxies, which makes them look more like
false S-warps than U-warps, the latter being more frequent 
for homogeneous disks.

\subsection{Simulated true warps}  

We also considered models with genuine warps, assuming a linear
slope for the plane, as soon as the radius is larger than some
critical value $r_{warp}$ = 8kpc. The slope corresponds to
an angle of $wa$ = 10$^\circ$, to obtain after
projection the order of magnitude of those observed
in the optical images. We varied the position angle $PA_w$ of
the warp line of node with the spiral. The outer parts mid-plane altitude
with respect to the main plane of the galaxy is then:
$$
<z> = (r-r_{warp}) tang(wa) cos (\theta - PA_w)
$$
We consider only straight line of nodes, which is justified
from the survey of Briggs (1990). The HI warps, that the 
optical warps tend to follow, have a straight line of nodes
from R$_{25}$ to $R_{26.5}$ (the Holmberg radius). 
After $R_{26.5}$, the line of nodes advances in the direction of
galaxy rotation, and therefore forms a leading spiral.    

\begin{figure}
\psfig{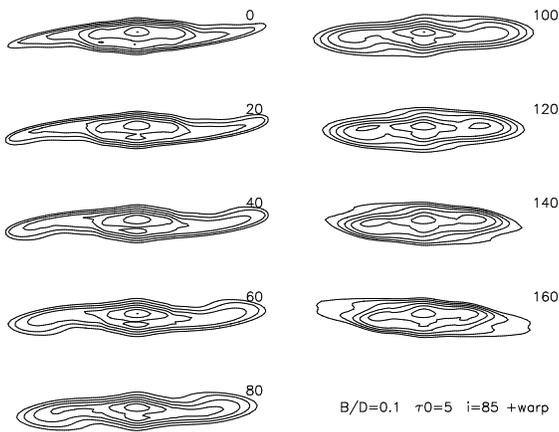}
\caption{ Some typical results from the genuine warp simulations: 
logarithmic contours
of inclined ($i= 85^\circ$) galaxies, with bulge-to-disk ratio $B/D = 
0.1$, and a total edge-on optical depth of $\tau_0=5$. 
The position angle $PA_w$ of the warp is here $0^\circ$.}
\label{b1t5i85w}
\end{figure}

\begin{figure}
\psfig{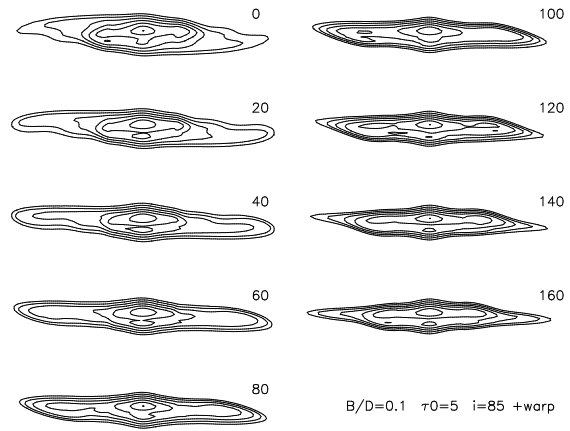}
\caption{ Same as previous figure, but now 
the position angle $PA_w$ of the warp is $90^\circ$.}
\label{b1t5i85wp}
\end{figure}

A sample of our galaxy models is plotted in fig \ref{b1t5i85w}
and \ref{b1t5i85wp}.
This allowed us to determine the number of genuine warps that 
are visible, compared to those that go un-detected because
their maximum height above the plane is along the line of sight.
 The latter cases suffer a bias against selection in a flat galaxy
sample, since the resulting effet is to thicken the apparent plane.
 From these models, and given the observed axis ratio distribution,
we estimate that we do not miss more than 
20\% of the warps through projection effects.

\section{Discussion and conclusion}  

We have simulated spiral projection effects cumulated to
dust-absorption effects on the appearance of warps 
in inclined galaxies. We find that indeed inclined $m=2$ spirals
can mimick warps, but the effect is not so serious for very
flat galaxies, at high inclination on the sky. We estimated that
no more than 15\% of S-shaped warps could be due to this phenomenon.
 We also emphasize that these projection effects could also 
subtract some genuine warps from the resulting apparent warp galaxies,
and therefore, the effect can somewhat be compensated.

From the simulations of real warped planes, we find that the fraction
of missed warps because of projection effects, i.e. those with line of
nodes perpendicular to the line of sight, is less than half, contrary
to what could be expected. This is due to the fact that a warp along
the line of sight thickens the apparent disk, and the resulting galaxy
is then not included in the sample of "flat" galaxies, as we have defined here.
We estimate that in our sample, we do not miss more than 20\% of
the true warps.

 Our observed statistics confirm the suspected influence of tidal interactions
in warping distortions. 
 The fact that large galaxies appear less warped in average might then 
be only due to the fact that they require more massive companions,
which are less frequent.
 The fact that most galaxies are warped, even in isolation, confirms
the persistence of the perturbation, even if triggered by a companion.
 
The percentage of warps is also to be compared to the percentage
of asymmetric galaxies.
 Richter \& Sancisi (1994) on a sample of 1700 galaxies observed in HI
deduced that at least 50\% of them show asymmetries, non-circularities,
lopsidedness. Zaritsky \& Rix (1997) on an optical sample find that
30\% of field spirals galaxies exhibit significant lopsidedness at large radii
(Zaritsky \& Rix 1997); this is attributed to companions.
 If warps could be attributed to companions, this will be
compatible with the fraction of warps that we find, provided that 
the warp is more long-lived than lopsidedness.

\acknowledgements{ We thank the referee, Roelof Bottema,
for his comments that have significantly improved the paper.
VR acknowledges support from Russian State Committee of
Higher Education, Russian Foundation for Basic Research (98-02-18178)
and from French Minist\`ere de la Recherche et de la Technologie.
}

\end{document}